\documentclass[12pt,a4paper,reqno]{article}
\usepackage{amsmath}
\usepackage{amssymb}
\usepackage{amsthm}
\usepackage{graphicx}
\usepackage[dvips]{color}

\newcommand{\blankline}{\vspace{\baselineskip}}

\renewcommand{\(}{\left(}
\renewcommand{\)}{\right)}

\newcommand{\e}{\mbox{e}}
\renewcommand{\hbar}{\hslash}

\newcommand{\ox}{\otimes}
\newcommand{\<}{\langle}
\renewcommand{\>}{\rangle}

\newcommand{\be}{\begin{equation}}
\newcommand{\ee}{\end{equation}}

\theoremstyle{plain} 

\theoremstyle{definition}

\theoremstyle{remark}

\begin{document}
\title{The Everett-Wheeler interpretation and the open future}
\author{Anthony Sudbery\\[10pt] \small Department of Mathematics,
    University of York,\\[-2pt] \small Heslington, York, England YO10 5DD\\ 
    \small  Email:  as2@york.ac.uk}
\date{Talk given at\\Advances in Quantum Theory\\[2pt]Linn\ae us University, V\"axj\"o, Sweden, 14 June 2010\\and at\\16th UK and European meeting on the Foundations of Physics\\Aberdeen, 5 July 2010}

\maketitle

\begin{abstract} I discuss the meaning of probability in the Everett-Wheeler
interpretation of quantum mechanics, together with the problem of
defining histories. To resolve these, I propose an understanding of
probability arising from a form of temporal logic: the probability of a future-tense proposition is identified with its truth value in a many-valued and context-dependent logic.  In short, probability is degree of truth. These ideas appear to
be new (though I expect correction on this), but they are natural
and intuitive, and relate to traditional naive ideas of time and
chance. Indeed, I argue that Everettian quantum mechanics is the
only form of scientific theory that truly incorporates the
perception that the future is open.

\end{abstract}

\newpage

\begin{center}
FOREWORD
\end{center}

This talk was delivered under the title ``The Everett interpretation and the open future". After the talk, a member of the audience expressed surprise that I had never mentioned many worlds or splitting observers. I explained that I regarded the ``many worlds interpretation'' as a betrayal of Everett's original idea: it introduces problems of defining worlds and moments of splitting which are equivalent to the measurement problem which Everett intended to solve (see \cite{QMPN} p. 221). Since then I have read Peter Byrne's biography of Everett, which makes it appear that I may have been wrong about this: according to Byrne, the subtlety which I see as the essence of Everett's original paper \cite{Everett} was imposed on him by his thesis adviser, John Wheeler, and what I regarded as extraneous features subsequently added by Bryce de Witt were in fact an essential part of Everett's vision. I have therefore reverted to an older nomenclature, now generally disapproved, and restored Wheeler's name to the interpretation I intend.

\begin{center}
INTRODUCTION
\end{center}

There are two serious problems for the Everett-Wheeler (EW) interpretation of quantum mechanics (and equally for the Everett or many-worlds interpretations, if we want to distinguish them). The first, which is generally acknowledged and has had many proposed solutions, is the place of probability in the theory. If all outcomes of an experiment actually occur, how can it make sense to talk about the probability that any one of them will occur? Yet probability is a central feature of quantum mechanics, and we had better include it if we want to be understood as talking about quantum mechanics at all.

The other problem, which particularly troubled John Bell \cite{Bell:cosmologists}, is that of the past and the future for an observer in an Everettian universe. Each of us has a sequence of experiences extending back in the past, and expects this sequence to continue into the future; but there is nothing in the EW interpretation to link any particular branch of the universal state vector at some past date with the branch that we inhabit today. Likewise I have no particular future. Clearly this problem is linked to the problem of probability; for if there is nothing to make a particular outcome true or false, how can we make sense of the probability that it is true?

In this talk I will propose, if not a solution, at least a way of approaching these two problems which renders them no special burden for Everettians. Probability and time were mysterious long before we had heard of quantum mechanics; I will suggest that they do not acquire any extra mystery in the EW interpretation.

\begin{center}
INCOMPATIBILITIES
\end{center}

Some of the differences between the EW and other interpretations of quantum mechanics are set out below.

\blankline

\begin{tabular}{lll}

$|\Psi\>$ is an objective property && $|\Psi\>$ represents a subjective \\
of the physical world              && state of knowledge\\
                                   && (Bayesians) \\[15pt]
$|\Psi\>$ evolves continuously     && $|\Psi\>$ collapses\\ 
by the Schr\"odinger equation      && after a measurement\\
                                   && (Copenhagen)\\[15pt]
$|\Psi\>$ describes a single system && $|\Psi\>$ describes an ensemble of systems

\end{tabular}

\blankline

It seems evident that the statements on the left contradict those on the right; in each pair of statements, both cannot be true. I want to suggest that these apparently contradictory statements can nevertheless be compatible, because different kinds of truth apply to the two columns. To explain this, let us consider --- what else? --- Schr\"odinger's cat experiment.

\begin{center}
SCHR\"ODINGER'S CAT
\end{center}

Schr\"odinger's sad story \cite{Schrcat} is often presented as a challenge to quantum mechanics. When the unfortunate cat has been in 
Schr\"odinger's diabolical device for a time $t$, the crude argument (not Schr\"odinger's!) goes, its state is
\[
|\psi_\text{cat}(t)\> = \e^{-\gamma t}|\text{alive}\> + \sqrt{1 - \e^{-2\gamma t}}|\text{dead}\>
\] 
So why don't we see such superpositions of live and dead cats?

The answer is simple. If we are watching the cat, hoping to see a superposition like the above, the interaction by which we see it actually produces the entangled state
\[ 
|\Psi(t)\> = \e^{-\gamma t}|\text{alive}\>_\text{cat}|\stackrel{\centerdot\;\centerdot}{\smile}\>_\text{observer} + \sqrt{1 - \e^{-2\gamma t}}|\text{dead}\>_\text{cat}|\stackrel{\centerdot\;\centerdot}{\frown}\>_\text{observer}
\]
in which $|\stackrel{\centerdot\;\centerdot}{\smile}\>$ is the observer state of seeing a live cat and $|\stackrel{\centerdot\;\centerdot}{\frown}\>$ is the state of seeing a dead cat. Nowhere in this total state is there an observer seeing a superposition of a live and a dead cat.

But that doesn't tell us what there actually \emph{is} in the state $|\Psi\>$. In order to understand the meaning of this superposition, it is helpful to look at it more carefully.

\begin{center}
WHAT IS TRUTH?
\end{center}

If the observer is watching the cat continuously over the period from time $0$ to time $t$, they will be able to note the time, if any, at which they see the cat die. Then the joint state of the cat and the observer is something like
\begin{multline}
|\Psi(t)\> = \e^{-\gamma t}|\text{alive}\>_\text{cat}|\text{``The cat is alive''}\>_\text{observer}\\ + \int_0^{t'}\e^{-\gamma t'}|\text{dead}\>_\text{cat}
|\text{``I saw the cat die at time $t'$''}\>_\text{observer}dt'
\end{multline}
in which the observer states contain propositions which are physically encoded in the brain of the observer. But what is their status as propositions; are they true or false?  Each is believed by a brain which has observed the fact it describes, and that fact belongs to reality. As a
human belief, each statement could not be more true. Yet they cannot all be true, for they contradict each other.

This conflict shows the necessity of considering the context in which a statement is made when discussing its truth value. When this is done, it becomes possible for contradictory statements to be simultaneously true, each in its own context. 

\begin{center}
COMPATIBILISM
\end{center}

In general, the state of the universe can be expanded in terms of the states of any observer inside the universe as
\[
|\Psi(t)\> = \sum_n|\eta_n\>|\Phi_n(t)\>
\]
where $|\eta_n\>$ are all possible experience states of the observer (an orthonormal basis of observer states) and $|\Phi_n(t)\>$ are corresponding states of the rest of the universe at time $t$. The actual observer can only experience being in one of the states $|\eta_n\>$, and it is true in this state that the only experience the observer has is $\eta_n$; the observer is justified, at time $t$, in deducing that the rest of the universe is in the unique state $|\Phi_n(t)\>$. This is the internal truth relative to the experience state $|\eta_n\>$.

But there is also the \emph{external} truth that the state of the whole universe is $|\Psi(t)\>$. From this standpoint all the experiences $\eta_n$ truly occur.\footnote{In the many-worlds interpretation it is said that they occur in different worlds which are all ``equally real'' \cite{Byrne}, though the justification for that ``equally'' is dubious; however, we there is no need to adopt that ontology.} Thus there are the following two types of truth involved.

{\bf External truth:} The truth about the universe is given by a state vector $|\Psi(t)\>$ in a Hilbert space $\mathcal{H}_U$. If this can be factorised as 
\[
\mathcal{H}_U = \mathcal{H}_S\ox\mathcal{H}_E
\]
where $\mathcal{H}_S$ contains states of an experiencing observer, then 
\[
|\Psi(t)\> = \sum_n|\eta_n\>|\Phi_n(t)\>
\]
and all the experiences $|\eta_n\>$ for which $|\Phi_n(t)\> \ne 0$ actually occur at time $t$.

\bigskip

{\bf Internal truth} from the perspective $|\eta_n\>$: I actually have experience $\eta_n$, which tells me that the rest of the universe is in the state $|\Phi_n(t)\>$. This is an objective fact; everybody agrees with me.

\begin{center}
EXTERNAL vs INTERNAL
\end{center}

This distinction is a special case of a general opposition which was introduced and carefully discussed by the philosopher Thomas Nagel \cite{Nagel:nowhere}. Scientists might be tempted to exalt the external statement as the objective truth, downgrading internal statements as merely subjective. Indeed, Nagel himself uses this terminology. However, he is at pains to emphasise that the truth of an internal statement has a vividness and immediacy, resulting from the fact that it is actually experienced, compared to which external truth is ``bleached-out''. This applies most obviously in the contexts of ethics and aesthetics, but we would do well to remember it in our scientific context; as I have pointed out above, it is the internal statement which has the scientific justification of being supported by evidence, and is objective in the usual sense that it is empirical and is agreed by all observers who can communicate with each other.

But the situation is more complicated than this might suggest. It is not that there is a God-like being who can survey the whole universe and make statements about the universal state vector, distinct from us physical beings who are trapped in one component of $|\Psi\>$. It is we physical beings who make statements about $|\Psi\>$, for good theoretical reasons, from our situation in which we experience just the one component $|\eta_n\>|\Phi_n\>$. From that perspective, what are we to make of the other components $|\eta_m\>|\Phi_m\>$?

Consider a measurement process, in which an initial state $|\Psi(0)\> = $ $|\eta_1\>|\Phi_1(0)\>$, containing only one experience $|\eta_1\>$, develops in time $t$ to an entangled state $|\Psi(t)\> = \sum|\eta_n\>|\Phi_n(t)\>$. The external statement is: 
\begin{quotation} $|\Psi\>$ represents a \emph{true} statement about the universe, and all its components are \emph{real}. \end{quotation}
The observer who experiences only $|\eta_n\>$ must say: 
\begin{quotation} I know that only $|\eta_n\>$ is \textbf{real} (because I experience only that), and therefore $|\Phi_n(t)\>$ represents a \textbf{true} statement about the rest of the universe. But I also know that $|\Psi(t)\>$ is \emph{true} (because I've calculated it). The other $|\eta_m\>|\Phi_m\>$ represent things that \textbf{might} have happened but \textbf{didn't}.
\end{quotation}
This statement is font-coded using bold type for internal (vivid, experienced) judgements, and italic for external (pale, theoretical) ones, even though these are made by an internal observer.

\begin{center}
COLLAPSE
\end{center}
 
The observer in this measurement process might go on to say:
\begin{quotation}
I \textbf{saw} a transition from $|\eta_1\>$ to $|\eta_n\>$ at some time $t' < t$. But I \emph{know} that $|\Psi(t')\>$ didn't collapse. The other $|\eta_m\>|\Phi_m\>$ \textbf{might} come back and interfere with me in the future; but this has very low \textbf{probability}.
\end{quotation}
This shows how the conflicting statements about time development in the EW and Copenhagen interpretations can after all be compatible. The continuous Schr\"odinger-equation evolution postulated by the EW interpretation refers to the external view; the collapses postulated by the Copenhagen interpretation refer to the internal view, i.e.\ to what we actually see. The occurrence of collapse becomes a theorem rather than a postulate (hopefully --- more on this later). We also see that it is only the internal statement that mentions probability. But what does it mean?

\begin{center}
WHAT IS PROBABILITY?
\end{center}

The meaning of probability is a long-standing philosophical problem (see \cite{Gillies}). It seems likely that there are in fact several distinct concepts which go by the name of probability, sharing only the fact that they obey the same mathematical axioms. The clearest of these, perhaps, is degree of belief, which has the advantage that it can be defined operationally: someone's degree of belief in a proposition is equal to the odds that they are prepared to offer in a bet that the proposition is true. The subjective nature of this concept seems to chime with the fact that it belongs in internal statements, as we have just seen, and indeed similar views of probability are often adopted by Everettians even though their general stance is objectivist.

However, we have also seen that ``internal'' should not be equated with ``subjective'', and our experience in a quantum-mechanical world seems to require a description in terms of objective chance. Things happen randomly, but with definite probabilities that cannot be reduced to our beliefs. The value of the half-life of uranium 238 is a fact about the world, not a mere consequence of someone's belief. 

Such objective probability can only refer to future events.

\begin{center}
PROBABILITY AND THE FUTURE
\end{center}

What kinds of statements can be made at time $t$ about some future time $s > t$, if the universal state vector is known to be $|\Psi(t)\>$ and its decomposition with respect to experience states of a particular observer is $\sum_n|\eta_n\>|\Phi_n(t)\>$? From the external perspective, the future state $|\Psi(s)\>$ is determined by the Schr\"odinger equation and there is no question of any probability. From the internal perspective relative to an experience state $|\eta_n\>$, there is a range of possible future states $|\eta_m\>$, and probabilities must enter into the statement of what the future state will be. But here is a fundamental problem: there is \emph{no such thing} as what the future state will be. As Bell pointed out, there is no connection between a component of $|\Psi\>$ at one time and any component at another time; so what is it that we can assign probabilities to? How can ``the probability that my state will be $|\eta_m\>$ tomorrow" mean anything when ``my state will be $|\eta_m\>$ tomorrow" has no meaning?

\begin{center}
THE CLASSICAL FUTURE
\end{center}

This puzzle takes us back to ways of thinking that are much older than quantum mechanics, indeed older than all of modern science. The success of Newtonian deterministic physics has led us to assume that there always is a definite future, and even when we drop determinism we tend to continue in the same assumption. There is a future, even if we do not and cannot know what it will be. But this was not what Aristotle believed, and maybe it is not what we believed when we were children. 

Aristotle, in a famous passage \cite{seabattle}, considered the proposition ``There will be a sea-battle tomorrow''.
He argued that this proposition is neither true nor false (otherwise we are forced into fatalism). Thus he rejected the law of excluded middle for future-tense statements, implying that they obey a many-valued logic. Modern logicians \cite{Prior} have considered the possibility of a third truth-value in addition to ``true" or ``false", namely $u$ for ``undetermined", for future-tense statements. But, interestingly, Aristotle admitted that the sea-battle might be more or less likely to take place. This suggests that the additional truth values needed for future-tense statements are not limited to one, $u$, but can be any real number between 0 and 1 and should be identified with the probability that the statement will come true. Turning this round gives us an objective form of probability which applies to future events, or to propositions in the future tense; in a slogan,
\[
\text{Probability} = \text{degree of truth}.
\]

\begin{center}
PROBABILITIES AS TRUTH VALUES
\end{center}

This translation of Aristotle's position seems so natural that it has surely been developed already. However, I have been unable to find it in the literature of probability theory, temporal logic or many-valued logic. In an early paper \cite{Luk:prob} (earlier than what are generally regarded as the first papers on many-valued logic) \L ukasiewicz introduced truth values between 0 and 1 and equated them with probabilities, but in a different sense from that of quantum mechanics (he was searching for a notion of objective probability, but found it only in propositions containing a free variable; he rejected any application to future-tense statements with no free variables, to which he thought probability did not apply because at this time (1913) he believed in determinism). The notion of ``degree of truth" occurs in fuzzy logic and philosophical discussions of vagueness (for a critical account see \cite{vagueness}), but not, as far as I can see, in the philosophy of probability. Maybe this is because it requires modifications to some of the usual properties of truth values. 

First, truth values are usually assumed to have the property that logical connectives like $\land$ (and) and $\lor$ (or) are "truth-functional", i.e.\ the truth values of $p\land q$ and $p\lor q$ are determined by those of $p$ and $q$. The probability of a compound statement, however, is constrained but not determined by the probabilities of its constituents: for the probabilities of $p\land q$ and $p\lor q$ we only have inequalities
\begin{align*}
P(p), P(q) &\ge P(p\land q) \ge 1 - P(p) - P(q),\\
P(p), P(q) &\le P(p\lor q) \le P(p) + P(q).
\end{align*}
Although degrees of truth for vague statements are usually assumed to be truth-functional, some authors have argued that they should behave like probabilities, as above. (\cite{Edgington}; see also \cite{vagueness}).

A more radical departure from the usual properties of truth values is that the probability of a proposition referring to a future time $s$ depends not only on the proposition itself (taking the time $s$ to be a part of the proposition), but also on the time $t$ at which the proposition is considered. For an observer experiencing the state $|\eta_n\>$ at time $t$, the probability of experiencing $\eta_m\>$ at a future time $s$ is 
\be\label{prob}
\frac{\left|\(\<\eta_m|\<\Phi_m(s)|\)\e^{-iH(s-t)/\hbar}\(|\eta_n\>|\Phi_n(t)\>\)\right|^2}{\<\Phi_m(s)|\Phi_m(s)\>\<\Phi_n(t)|\Phi_n(t)\>}.
\ee
Thus our account of probability requires that the truth value of the proposition ``My experience at time $s$ will be $|\eta_m\>$" depends on the time at which the proposition is uttered. Such dependence on a context of utterance is nothing new (consider the truth value of the proposition ``it is raining"), but that can usually be understood as being because the context of utterance is needed to fill in an incomplete proposition (in our example, it provides the time and place at which it is raining). There is no question of this in the case of the occurrence of $|\eta_m\>$ at time $s$.

\begin{center}
THE TRUTH OF THE PAST
\end{center}
The classical position (in classical philosophy, if not classical physics) would be that propositions referring to the past and present are either true or false; it is only the future that is uncertain. Thus if $P(s, t)$ is the truth value at $t$ of a proposition referring to time $s$, we should have
\begin{align}\label{past}
0 \le P(s, t) \le 1 &\quad \text{ if } s > t,\notag\\
P(s,t) = 0 \text{ or } 1 &\quad \text{ if } s \le t.
\end{align}
It is not clear whether this can or should be maintained in quantum theory. Bell's point \cite{Bell:cosmologists} about the lack of connection between experience states at different times applies to the past as much as the future. We do have an empirical warrant for our past states, as we do not for future states, in the form of memory, which is not symmetric under time reversal. This has yet to be modelled quantum-mechanically, but hopefully it can be shown that there is a physical process which leaves a record in the state at one time of a sequence of states in the past, and that this record is consistent with the probabilities \eqref{prob}. Maybe it would go further and establish transition probabilities such as have been postulated for modal interpretations \cite{BacciaDickson, verdammt}. If so, this might provide justification for truth values satisfying \eqref{past}. The theory would still be vulnerable to Bell's charge of temporal solipsism --- a memory state is still a present state, and does not constitute a genuine past --- but this would  be open to debate. 

\newpage

\begin{center}
THE OPEN FUTURE
\end{center}

We find it hard to accommodate the idea that there is no definite future in a scientific theory. To be sure, we have indeterministic theories in which the future is not uniquely determined by the past, but such stochastic theories deal with complete histories encompassing past, present and future; probabilities refer to which of these histories is actualised. Indeterminism consists of the fact that there are many such histories containing a given past up to a certain time, so the future extension is not unique; but the underlying assumption is that only one of these future histories is real, so that the future is fixed even though it is not determined. In contrast, the EW formulation of quantum mechanics (or, as Bell \cite{Bell:cosmologists} calls it, the ``Everett (?) theory") is the only scientific theory in which the future is open. This is not a problem for the theory; it tells a truth which we should recognise.


%
%


\end{document}